\documentclass{llncs}

\usepackage[linesnumbered]{algorithm2e}
\usepackage{graphicx}
\usepackage{textcomp}
\usepackage{enumitem}
\usepackage{url}
\usepackage{floatrow}
\usepackage{booktabs}
\usepackage{multirow}

\usepackage[nodisplayskipstretch]{setspace}
\AtBeginDocument{%
  \addtolength\abovedisplayskip{-0.5\baselineskip}%
  \addtolength\belowdisplayskip{-0.5\baselineskip}%
}

\newcommand{\Comment}[1]{}

\begin{document}

\title{Architecting Dependable Learning-enabled Autonomous Systems: A Survey}

\author{
Chih-Hong Cheng\inst{1}
\and
Dhiraj Gulati\inst{1}
\and
Rongjie Yan\inst{2}
}

\institute{
	fortiss - Research Institute of the Free State of Bavaria, Germany \\
	\and
	State Key Laboratory of Computer Science, China \\
    \vspace{1mm}
    \texttt{\{cheng,gulati\}@fortiss.org, yrj@ios.ac.cn}\\
}

\maketitle

\vspace{-5mm}

\begin{abstract}

We provide a summary over architectural approaches that can be used to construct dependable learning-enabled autonomous systems, with a focus on automated driving. We consider three technology pillars for architecting dependable autonomy, namely \emph{diverse redundancy}, \emph{information fusion}, and \emph{runtime monitoring}. For learning-enabled components, we additionally summarize recent architectural approaches to increase the dependability beyond standard convolutional neural networks. We conclude the study with a list of promising research directions addressing the challenges of existing approaches.

\end{abstract}

\section{Introduction}

Recent fatalities of autonomous vehicles on public roads such as the  2018 Uber crash~\cite{UberCrash} have raised the necessity of creating safe autonomous systems. 
Apart from  having a rigorous verification and validation (V\&V) plan,  automotive safety standards such as ISO~26262 or ISO/PSA~21448 also emphasize the importance of  \emph{designing dependable system architectures}. Although researchers have developed various techniques that may act as building blocks for dependability, what remains missing, however, is a systematic categorization over existing research results while mapping these techniques to a reference architecture for dependable autonomy.

In this paper, based on an extensive review of prior works, we provide a summary over architectural approaches that can be used to construct dependable autonomous systems, with a focus on automated driving. We consider \emph{diverse redundancy}, \emph{information fusion}, and \emph{runtime monitoring} as key architectural means for building dependable autonomous systems.  

\begin{itemize}
    \item \emph{Diverse redundancy} refers to the use of different paradigms or algorithms to achieve the end-to-end workflow from sensing to actuation. 
    
    \item \emph{Information fusion} refers to the technique of merging homogeneous or heterogeneous information from  diverse sources into one.
    
    \item \emph{Runtime monitoring} refers to the technique of detecting obvious or intriguing abnormality of information being produced by components or subsystems. 
\end{itemize}

Data-driven engineering (learning from data) is increasingly  used to create perception, prediction, or even decision components whose precise specification is hard to characterize in automated driving tasks. Therefore, as an addition to the above list, we also include 

\begin{itemize}
    \item \emph{architecture approaches for dependable machine learning~(ML)} that go beyond standard convolutional neural networks (CNN). 
\end{itemize}

\begin{figure}[t]
    \centering
    \includegraphics[width=0.8\textwidth]{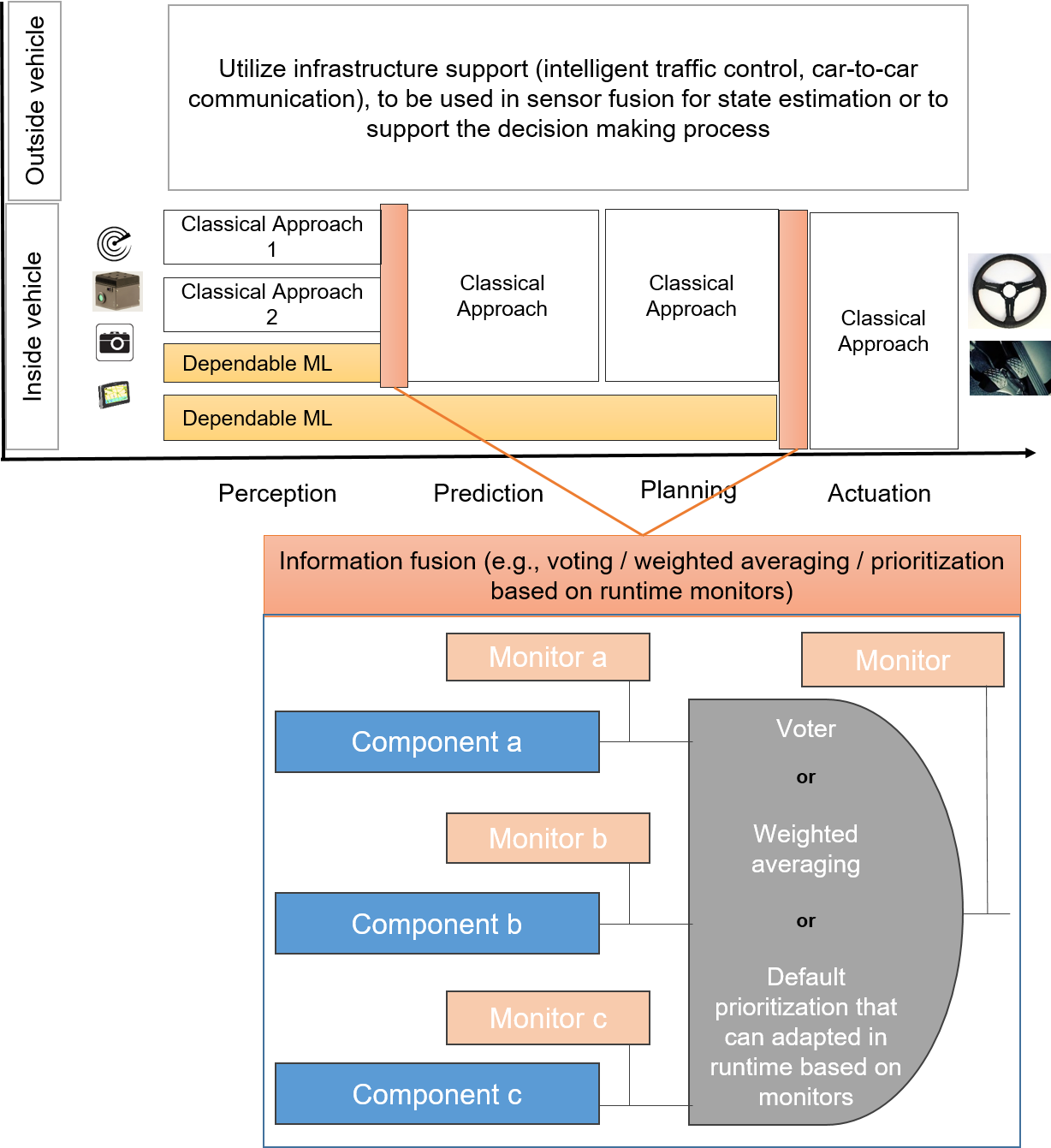}
    \caption{High-level illustration how architectural diversity, information fusion, monitoring, and dependable ML components contribute to the overall dependable architecture.}
    \label{fig:architecture}
    \vspace{-2mm}
\end{figure}

How the above mentioned four ingredients constitute to dependable architectures for autonomous driving can be understood using the example in Figure~\ref{fig:architecture}. It can be seen that a dependable ML is crucial to ensure the quality of the subsystems learned from data, runtime monitoring examines erroneous behaviors of components created with diversity, and lastly, information fusion techniques may adjust the decision based on the result of monitors.

\section{Architecting Dependable Autonomous Driving Systems}

\begin{figure}[t]
    \centering
    \includegraphics[width=\textwidth, trim=0.9cm 5.0cm 1.5cm 3.5cm, clip]{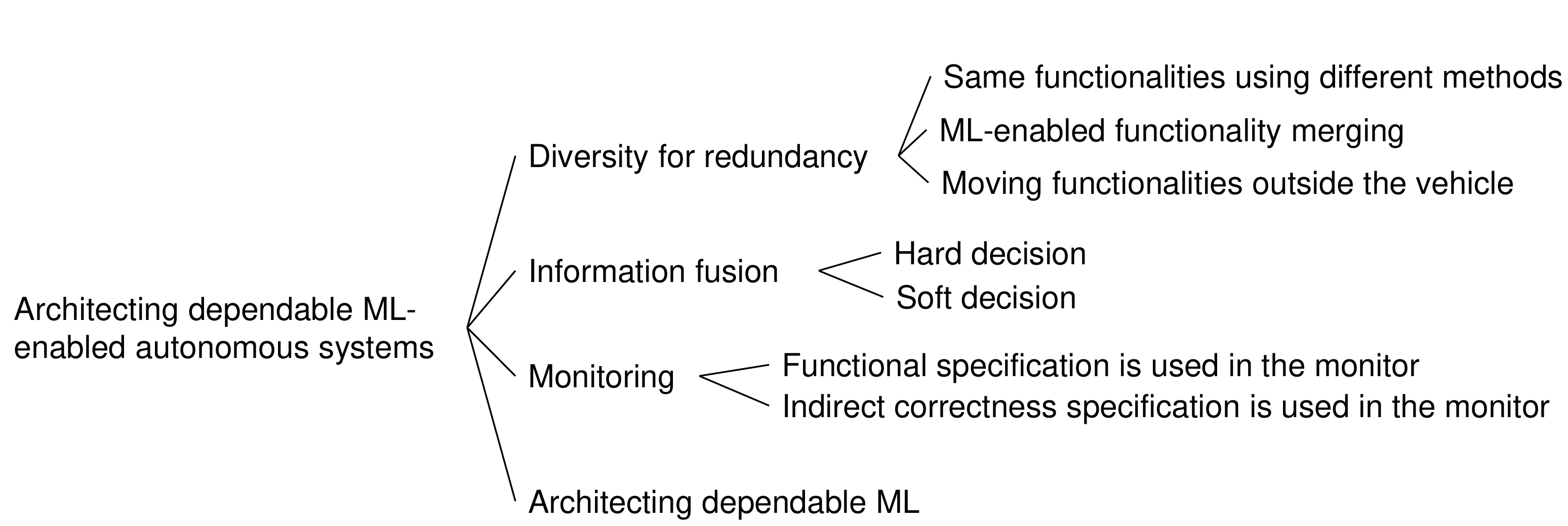}
    \caption{High-level categorization of the review structure.}
    \label{fig:categorization}
    
\end{figure}

 Figure~\ref{fig:categorization} outlines the broad categorization of the four points mentioned in the previous section. The state-of-the-art architectures are further explored and divided based on this categorization.

\subsection{Diverse redundancy}

Intuitively speaking, realizing autonomous driving can be understood as implementing a transducer (function with states) that takes high-dimensional input (e.g., vision) and derives output in low-dimensional space (steering, acceleration and break). The overall process can be further refined into a sequence of high-level functionalities including \emph{perception}, \emph{prediction}, \emph{planning}, and \emph{actuation}. From the related work, there are three possible means to achieve diversity:

\begin{enumerate}

\item{\bf Implementing the same functionality using different methods.} For each of the high-level functionalities, one can implement the same functionality using different methods. 

\begin{itemize}
    \item For perception, one can either use stereo vision or lidar to perform 3D scene reconstruction, which internally employs different algorithms. Vision-based perception techniques involve many dimensions, such as distance estimation~\cite{yebes2015visual}, painted line detection~\cite{olivares2016vision}, and 3D dense semantic map~\cite{yang2018large}. Methods of perception with lidar have shifted from classification~\cite{brodu20123d} to detection with convolutional networks, such as vehicle detection~\cite{li2016vehicle}, zone detection~\cite{maturana20153d}, and  semantic segmentation with conditional random field~\cite{wu2018squeezeseg}.  There is also a trend to fuse various sensory data for various detection purposes~\cite{berrio2017fusing,wulff2018early}.    Additional to image perception, the work in \cite{DBLP:conf/icra/MarchegianiP17} could spot acoustic events, such as horns for smart vehicles, with anomaly detection for the urban soundscape model.
    \item For prediction, algorithm diversity includes probabilistic approaches, rule-based approaches, and ML-based approaches. For example,  using these information and chassis information, probabilistic prediction in~\cite{chae2018design} is conducted for ego vehicle and surrounding vehicle separately, to facilitate motion planning.  Using goal-directed planning, pedestrians can be predicted based on  common behavior patterns learned by a fully convolutional network operating on maps of the environment~\cite{DBLP:conf/icra/QianYW018}. By modeling the interaction of all its surrounding vehicles' trajectories, without over-the-air communication between vehicles, the method proposed in~\cite{dong2018continuous} could predict the continuous lane-change trajectory of a target car.  The work in~\cite{DBLP:conf/icra/HoermannBD18} combines  a Bayesian filtering technique for environment representation, and machine learning as long-term predictor, and applies future and past estimates to separate static and dynamic regions. 

    \item For planning and decision making, algorithm diversity includes sampling-based approaches, grid-based approaches, potential fields approach and others. There are several reviews for planning, such as~\cite{gonzalez2016review,schwarting2018planning}. As summarized in \cite{schwarting2018planning}, planning with input space discretization, such as lattice planners~\cite{pivtoraiko2009differentially} and road-aligned primitives~\cite{werling2012optimal} is simple and effective. Randomized planning, such as rapidly exploring random trees (RRT)~\cite{karaman2011sampling}, could explore large state space with a high computational cost.  Initially applied for path following, constrained optimization~\cite{liniger2015optimization} can also compute collision-free trajectories to avoid other traffic participants. The optimization scheme can also improve the quality of the solution~\cite{dolgov2010path}.  Considering the dynamic interaction between autonomous vehicles and human drivers, Sadigh et al. provide an optimization-based planner in the dynamical system~\cite{sadigh2016planning}. The recent review~\cite{schwarting2018planning} has further summarized results on planning and decision-making for autonomous vehicles.

    \item For actuation, algorithm diversity ranges from Proportional-Integral-Differential (PID) control to complex model-predictive control. For example, an embedded fuzzy-logic-based control system controls both speed and steering of  automated vehicles~\cite{naranjo2007using}. Considering the tire force saturations, a nonlinear feedback strategy is proposed to improve the transient performance and eliminate the steady-state errors in path-following control~\cite{wang2016composite}.

\end{itemize}

\begin{figure}[t]
    \centering
    \includegraphics[width=0.8\textwidth]{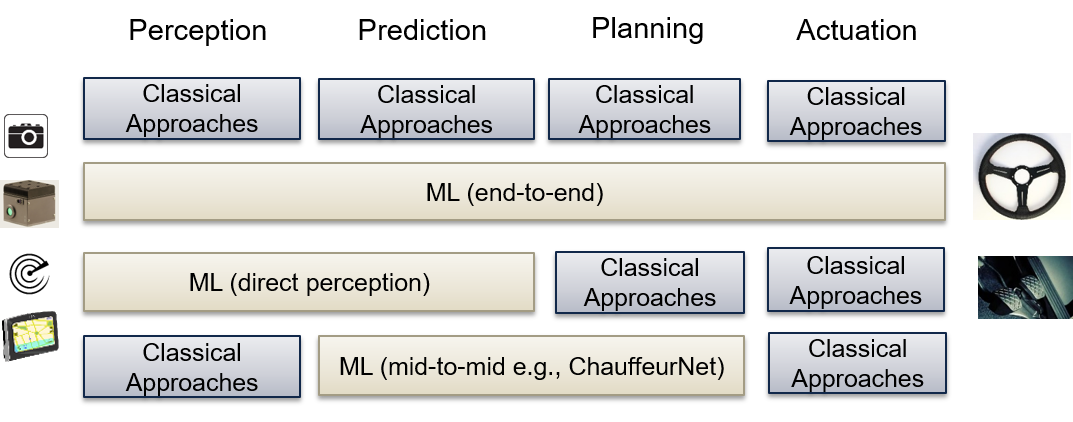}
    \caption{Creating architectural diversities by  merging functionalities in the pipeline.}
    \label{fig:pipeline.diversity}
\end{figure}

\item{\bf Merging functionalities in the pipeline.}  Apart from the classical pipeline and modular approach, one architectural diversity is to merge  functionalities by utilizing machine learning. Figure~\ref{fig:pipeline.diversity} illustrates the possible architectures using this paradigm. 

\begin{itemize}
    \item One extreme is an end-to-end approach (behaviour reflex)~\cite{pomerleau1989alvinn,xu2017end,hecker2018end,koutnik2013evolving}, which directly trains a neural network as a single transducer (i.e., it merges four functionalities into one). To achieve the goal, it takes an image and other sensory inputs and directly outputs the steering angle and acceleration parameters. 
    
    \item While end-to-end approaches merges four functionalities into one, diversities can also be achieved by building a neural network that only performs the task of a subset of functionalities, i.e., the so called \emph{mid-to-mid} approach.
    
    \begin{itemize}
        \item  One realization is \emph{direct perception}~\cite{chen2015deepdriving,sauer2018conditional} which uses machine learning to translate the high dimensional sensor input to high-level features called affordances (e.g., distance to the preceding car, or detection of traffic lights), such that the features are used as input for the vehicle planning and control algorithms. 
        \item Another realization can be found in the work of  \emph{ChauffeurNet}~\cite{bansal2018chauffeurnet}. It starts by assuming that the bounding boxes of objects are made available on the bird's view, and the system trains an RNN to perform motion prediction and path planning. 
        
        \item  Yet another possibility is to go from (visual) perception to planning. The work by Song et al.~\cite{song2018learning} trains a motion planning system, where internally it combines a convolution neural network and a long short-term memory (LSTM) to extract spatial and temporary features of the input images.
    \end{itemize}
\end{itemize}

\item{\bf Moving intelligence outside the vehicle.} Apart from making a vehicle intelligent, the third diversity is to offload the task of the vehicle by preparing a digital infrastructure that facilitates vehicle-to-vehicle (V2V) or vehicle-to-infrastructure (V2I) communication \cite{hobertv2x}. For example, for a vehicle sensing a traffic light, state-of-the-art approaches  rely on visual perception. Alternatively
  with V2I communication support, the vehicle can also confirm the detection of traffic light by receiving the signal from the infrastructure. With these two methods a redundancy in functionality is achieved with support from intelligence outside the vehicle. Various work in the literature provides the benefits of using such an approach.

\begin{itemize}
    \item  (V2V) For highway scenarios, when the autonomous vehicle receives information from other vehicles or sensory data from the infrastructure, due to there being more data sources over the nearby vehicles, the precision can be higher as demonstrated in cooperative localization~\cite{kurazume1994cooperative}, a concept that dated in early~90s. The CoPark method allows negotiation between various vehicles within its V2V communication range and decides on the search areas for parking~\cite{aliedani2016investigating}.  Protocol MoZoa provides  a moving-zone based architecture to facilitate information dissemination, where  vehicles can collaborate with one another to form dynamic moving zones~\cite{Lin2017MoZoAM}.  
    \item  (V2I) As a recent example, the  research project Providentia~\cite{hinz2017proactive} builds a digital infrastructure on the German A9 highway to enable high speed V2I communication via 5G networks. For scenarios such as automated valet parking, the vehicle can retrieve information of free parking spaces by accessing the parking area infrastructure with occupancy detection camera~\cite{loper2013automated}. 
\end{itemize}

\end{enumerate}

\subsection{Information fusion}
 As already discussed, to develop a dependable architecture, redundancy is required and hence multiple such pipelines are executed in parallel. Therefore, a dependable autonomous driving system should fuse not only the sensor data but also data originating in each pipeline with each other at the same stage.

As seen in the architectures investigated in the last section, multiple components can be assembled in various combinations to generate distinct pipelines. It is then the job of information fusion to choose the best possible outcome and deliver it to the next component in the pipeline. Such fusion is also known as \emph{decision fusion}.

Table \ref{tab:ifsummary} (derived from \cite{liggins2017handbook} Chapter 5) presents an overview of the common methods used for the information to arrive at decisions.
\begin{table}[bh]
\caption{Common information fusion combination alternatives to make decisions}
\label{tab:ifsummary}
\noindent
\centering
\begin{tabular}{c c c } \toprule
{\textbf{Decision Type}} &
{\textbf{Method}} &
{\textbf{Description}}\\
\cmidrule(lr){1-3}
\multirow{3}{*}{\parbox{2.2cm}{Hard decision}} & {\parbox{3.5cm}{Boolean}} & {\parbox{6cm}{Apply logical AND, OR to combine independent values}} \\
\cmidrule(lr){2-3}
                               & {\parbox{3.5cm}{Weighted Sum Score}} & {\parbox{6cm}{Weight values by inverse of their errors and sum to derive score function}} \\
                               \cmidrule(lr){2-3}
                               & {\parbox{3.5cm}{M-of-N}} & {\parbox{6cm}{Confirm information based on m-out-of-n sources that agree}} \\
\cmidrule(lr){1-3}
\multirow{3}{*}{\parbox{2.2cm}{Soft decision}} & {\parbox{3.5cm}{Bayesian}} & {\parbox{6cm}{Apply Bayes rule to combine independent conditional probabilities}} \\
\cmidrule(lr){2-3}
                               & {\parbox{3.5cm}{Dempster-Shafer}} & {\parbox{6cm}{Apply Dempster's rule of combination to combine information belief functions}} \\
                               \cmidrule(lr){2-3}
                               & {\parbox{3.5cm}{Fuzzy variable}} & {\parbox{6cm}{Combine fuzzy variables using fuzzy logic (AND,OR) to derive combined membership function}} \\
\bottomrule
\end{tabular}
\end{table}
\emph{Hard decisions} constitute the fusion methods when they result in a single optimal choice. The \emph{soft decisions} are the result of methods where there may be more than one decision but each decision will have an uncertainty associated with it. Based on the requirements, different information fusion methods can be used in different parts of the automated driving pipeline.

\emph{Boolean information fusion} method is the easiest to understand and implement. If one pipeline reports an object and other does not, a Boolean OR decision will pass its existence to the next level. On the other hand, a Boolean AND will not report it. A combination of such operators based on multiple data points generated from one step in a pipeline with another pipeline can result in a simple yet powerful decision.

\emph{Weighted sum score} (also known as \emph{weighted sum model})~\cite{Triantaphyllou2000} is another method which can also be used to arrive at hard decisions. If multiple sources result in information in same units, this method can be used to combine and arrive at one concrete answer. Each information can also be assigned some weight (based on some past knowledge like error, past observations etc.) and then can be combined to arrive at a decision.

\emph{M-of-N} is a standard voting method where majority value is considered as the final decision. This methodology has been used in designing fault tolerant systems \cite{faulttolerant}. These methods can also be combined together with other to arrive at more powerful methods. For example, weights may be added before voting.

\emph{Bayesian decision making} is one of the most common soft decision making processes. The commonly seen Kalman Filters \cite{welch1995introduction} and Particle Filters \cite{arulampalam2002tutorial} use Bayesian decision making to arrive at final decisions with covariances matrices. The matrices represent the uncertainty of the arrived final decision.

The \emph{Dempster-Shafer theory}, also known as the theory of belief functions, is a generalization of the Bayesian theory of subjective probability~\cite{dempster1968generalization}.  Whereas the Bayesian theory requires probabilities for each information of interest, belief functions allow us to form degrees of belief for one information based on probabilities for a related information. Further, Dempster's rule is used for combining such degrees of belief when they are based on independent items of evidences~\cite{shafer1976mathematical}. Where Bayesian decision making relies on \textit{degree of agreement}, Demspter-Shafer method tries to measure \textit{absence of conflict}. The output is the decision with a belief associated with it. 

\emph{Fuzzy logic system} is a nonlinear mapping of an input information vector into a scalar output. Fuzzy set theory and fuzzy logic establish the specifics of the nonlinear mapping \cite{fuzzymendel}. Results from this non-linear mapping can be used along with Boolean operators to make decisions. Such rules which facilitate the fusion are called \emph{fuzzy rules}. The result then is defuzzied to arrive at the final output. 
Since the initial error is assumed to be fuzzy, the output error also remains fuzzy~\cite{fuzzyvarshney}.

The above methods by no means represent an exhaustive list of information fusion.
In fact more than one of the above approaches can be combined to result in \textit{Hybrid decisions}. For example, Bayesian approaches like Kalman Filter can be used to arrive at some information along with the required covariance matrix in multiple parallel pipelines. This matrix can be inverted and summed to derive appropriate weights for each of the pipeline process. For the next step a weighted sum score can be used to make the final decision.
Here we converted the soft decision to the hard decision. Additionally, this implies that the weights of contribution of each pipeline may change over time. 

\subsection{Monitoring / Runtime verification}

Monitors observe and report the abnormal outputs that deviate from the correct behavior by either observing the states or by analyzing finite history traces. As monitors are part of the concept of fault detection, isolation and recovery (FDIR)~\cite{hwang2010survey} which is seen in realizing safety critical systems such as avionics, here we restrict our focus to monitoring machine learning components and examining behavior of the autonomous vehicles. As detailed below, our categorization is based on whether a functional specification or an indirect correctness specification is used in the monitor.

\subsubsection{Monitoring indirect correctness information} For vision-based perception components, it is very hard to create correctness specification from first principles. Therefore, the labelled data is commonly considered the specification and neural networks are used for implementation. Due to the lack of specification, the goal of monitoring is to indirectly check if the \emph{data received in operation is not ``distant'' from training data in the input space} (as this can imply excessive extrapolation from training data), or if there exists \emph{distribution shift}, i.e., the data distribution  in operation significantly deviates from the data distribution in the training (as this can imply that the network is not trained properly).

\begin{enumerate}
\item {\bf Monitoring computed values of a neural network}. This line of research targets to detect if value computed by a neural network significantly deviates from the known values computed using the training data. Neural networks implementing \emph{zero-shot learning}~\cite{lampert2014attribute} can detect new classes (as abnormalities). The basic principle is to associate each class in training with a particular combination of predefined attributes\footnote{As an example, a stop sign in Germany can be associated with attributes such as ``red'', ``circle'', ``octagon''.}. So the monitor records each class with the unique 
	attribute pair. Whenever an input leads to a new set of attributes, the system considers it as a new type of object, and existing classification is not used.  
	The work of Cheng et al.~\cite{DBLP:journals/corr/abs-1809-06573} uses a monitor to store the neuron on-off activation patterns of training data as a set using binary decision diagrams (BDD)~\cite{bryant1992symbolic}. In operation, the monitor rejects the decision made by the network if the computed activation pattern is not contained in the BDD. 
	
	\item {\bf Monitor the distribution of predicted classes}. The goal is to understand if the distribution of the data has shifted. If so, then it might imply that the neural network originally trained under the specific distribution may not be able to perform (as there is a difference between training data set and operation data set) and the decision may be problematic. The common technique is based on a concept of \emph{change-point detection}, where one takes $k$ most recent points and compare it with $2k$ most recent points, i.e., to examine it using two bags of data~\cite{gretton2012kernel}.

\end{enumerate}

\subsubsection{Monitoring with precise \& correct specification} 

Specification-based monitors usually adopt formal specifications extended from LTL such as MTL~\cite{kane2015case}, STL~\cite{DBLP:conf/dac/WatanabeKLS18} or other extensions~\cite{DBLP:conf/rv/DokhanchiADF18}. These monitors are used to detect abnormal temporal behavior or assumption violation of individual modules~\cite{selyunin2016applying,DBLP:conf/rv/DokhanchiADF18} or between the interaction of various modules~\cite{DBLP:conf/dac/WatanabeKLS18}. The following categories are distinguished from  most of the specification-based monitors.

\begin{enumerate}
	\item {\bf Real-time monitors}. They are implemented on physical devices that are connected to the interested system. They are constrained by the frequency of the target, and the availability of resources~\cite{nguyen2016harmonia,selyunin2016applying}. Both qualitative~\cite{jakvsic2015signal} and quantitative~\cite{jakvsic2016quantitative} semantics for STL with real time operators and numerical inputs can be implemented in FPGA. 
	
	\item {\bf Failure prevention monitors}. Instead of purely reporting property or assumption violations, such monitors at runtime could enforce the system to behave correctly~\cite{DBLP:conf/dac/WuZWY17} or recover from abnormal behaviours~\cite{selyunin2017runtime}. Such monitors are automata-based, which are synthesized~\cite{DBLP:conf/dac/WuZWY17} or transformed~\cite{selyunin2017runtime} with respect to safety properties. Another type of monitors is to add interventions in the case of potential violations such that the system could reach functional states~\cite{masson2018tuning}.
	
	\item {\bf Falsification methods}. Instead of passively observing abnormal behavior from runtime traces, the methods attempt to find counterexamples for property or assumption violations with the given model of the system (or the actual system). For systems  containing machine-learning component, compositional methods are adopted to isolate possible falsifications and reduce the search space~\cite{dreossi2017compositional}. Subsequently, a temporal logic falsifier and a ML-based analyzer for detecting misclassifications cooperate to find falsifying executions of the considered model. 
	
\end{enumerate}

\subsection{Beyond standard CNN for dependable ML}

This section summarizes additional tactics that can be viewed as extending existing standard CNN engineering approaches with the goal of increasing dependability. Some of the described techniques are recent in that no application in autonomous driving has been reported. 

\begin{enumerate}
    \item {\bf Monte-Carlo Dropout Techniques~\cite{gal2016dropout}.}  Dropout is commonly used in \emph{training} to prevent overfitting of neural networks. The basic principle of dropout is to randomly disable some neurons of a neural network in training time, in order to make sure that the decision of a neural network is not tightly bounded to certain neurons. Contrarily, the Monte-Carlo Drop Out technique is a way of using dropout in \emph{operation time}. Let the original set of neurons be~$N$. Whenever dropout is performed first on the set~$A$ of neurons followed by set~$B$ of neurons, then~$N\setminus A$ and~$N\setminus B$ are essentially two different networks. A direct consequence is that one can perform drop out many times, in order to perform voting (thus it can be viewed as a special form of ensemble), or to perform uncertainty estimation – if out of $100$ networks with neurons being dropped out, $10$ of them have demonstrated deviating behaviours, then one may estimate that the network outputs the value with certainty $\frac{(100-10)}{100}= 90\%$. 
    
    \item {\bf Interpretable CNN~\cite{zhang2018interpretable,zhang2018interpreting}.} This division of research work focuses on modifying the training objective in design time, such that the loss function encourages (1) a neuron to be exclusively activated by a certain category~$C$ and to keep silent on other categories, and (2) a neuron to only be activated by a single region of the feature map, instead of being repetitively triggered at different locations. Whenever a network is constructed using the above approach~\cite{zhang2018interpretable}, as the activation of a neuron is specific to a location and to a specific type, it is then further possible to construct a decision tree by considering what forms a decision of classifying an object with category~$C$~\cite{zhang2018interpreting}.
    
    \item {\bf Neural networks with provable defences~\cite{kolter2017provable,steinhardt2017certified}.} The neural network is first trained using standard approaches. Subsequently, one performs techniques similar to \emph{robust optimization techniques}~\cite{ben2009robust} to the training of the neural network, which guarantees that any point that is sufficiently close to the training data demonstrates the same behavior like the training data. 
    
    \item {\bf Sequence models.} Unlike stateless CNN, recurrent neural networks such as LSTM~\cite{hochreiter1997long} or GRU~\cite{cho2014properties} can use their internal state (memory) to process a sequence of inputs. This allows ML models to incorporate temporal information to reduce noise and to make better predictions. Applications in autonomous driving include trajectory generation of ego vehicles~\cite{bansal2018chauffeurnet,altche2017lstm} and the prediction of surrounding objects~\cite{park2018sequence,zyner2018naturalistic}.
\end{enumerate}

\vspace{-4mm}

\section{Concluding Remarks}

\vspace{-2mm}

In this paper, we provide a review on architectural means  that can be used to construct dependable autonomous systems, with a focus on diverse redundancy, information fusion, runtime monitoring, and dependable ML techniques. The motivation of the study is to present researchers and practitioners a thorough understanding of the spectrum of existing approaches and to allow them to adapt the approaches in building dependable autonomous driving solutions.   
We conclude this report by highlighting some prominent research questions.

\vspace{-2mm}

\subsubsection{(Diverse redundancy)} The first  challenge for diverse redundancy refers to a need of a systematic and rigorous argument where in the operational design domain (ODD), implementations for diverse redundancy should not encounter a \emph{common failure scenario} where all implementations fails. This requires a careful verification and validation plan which is beyond the scope of this paper, but  many fruitful results are already targeting the V\&V aspects of autonomous vehicles~\cite{huang2018safety,hutchison2018robustness}. The second challenge refers to the use of ML for merging functionalities in the pipeline. For automotive domain,  there are different vehicle configurations due to shared vehicle platform and due to faster cycles to produce new generations, and arguably the ML function \emph{may not be directly transferrable} from one configuration to another. 

\vspace{-3mm}

\subsubsection{(Information fusion)} To arrive at correct decisions between parallel pipelines, first challenge is the \emph{fidelity of various models} (like process and error) used in the fusion. Most solutions assume linear gaussian data and noises and any deviation in this assumption gives incorrect results. Second challenge is the \emph{rigid fusion architectures}, where failure in one data source results in failure of the complete fusion process.
One method to overcome this is to provide feedback from the monitors to the fusion process. This would result in an \emph{adaptive fusion process}. This itself is a topic for ongoing research~\cite{hongadaptive,adaptivefusion}. There has been some success in direction of adaptive kalman filters~\cite{adaptivekalman,taftiadaptive,5158254}.

\vspace{-3mm}
\subsubsection{(Runtime monitoring)}

Apart from classical performance considerations, monitors for ML components may be improved by considering the following research questions. (1) How to better \emph{formulate the ``distance'' concept} for data in operation to data in the training set, such that it leads to the creation of resource efficient monitors with high error detection rate. (2) How to generate \emph{human interpretable monitors} for ML-components. Examples for interpretable monitors in object detection include ``impossibility for overly large objects to exist in the image''. For specification-based monitors, the foreseeable challenges are  (3) how to \emph{prioritize specifications to be monitored} based on a scientific characterization of occurrence, effect, and controllability, and (4) how to detect abnormality for \emph{unknown unknowns}, preferably with ML-based techniques. One final  challenge is to go from runtime monitoring to \emph{runtime enforcement}~\cite{bloem2015shield,wu2017safety}.

\vspace{-2mm}

\begin{spacing}{0.95}
\bibliographystyle{abbrv}

\end{spacing}

\end{document}